# Problem Complexity Research from Energy Perspective

*PAN Feng[1], ZHANG Heng-liang[2], Qi Jie[1]*

(1. School of Information Science and Technology, Donghua University, Shanghai 201620, China;
2. School of Power and Mechanical Engineering, Wuhan University, Wuhan 430072, China)

**Abstract**: Computational complexity is a particularly important objective. The idea of Landauer principle was extended through mapping three classic problems (sorting、ordered searching and max of *N* unordered numbers) into Maxwell demon thought experiment in this paper. The problems' complexity is defined on the entropy basis and the minimum energy required to solve them are rigorous deduced from the perspective of energy (entropy) and the second law of thermodynamics. Then the theoretical energy consumed by real program and basic operators of classical computer are both analyzed, the time complexity lower bounds of three problems' all possible algorithms are derived in this way. The lower bound is also deduced for the two $n \times n$ matrix multiplication problem. In the end, the reason why reversible computation is impossible and the possibility of super-linear energy consumption capacity which may be the power behind quantum computation are discussed, a conjecture is proposed which may prove **NP！=P**. The study will bring fresh and profound understanding of computation complexity.

## Ⅰ. INTRODUCTION

Certain amount of resources is necessary in order to solve any given problems. For instance, space (memory), time and energy are all essential to run an algorithm on a computer. Computational complexity is the study of the resources required to solve computational problems. Sometimes it seems immediately obvious if one problem is easier to solve than another. For instance, it's generally believed that to add two numbers is easier than to multiple them[1] (We'll see this is probably not true.). But in some cases, it may be very difficult to evaluate and measure the complexity of a problem. An important task of computational complexity is to find the minimum resources required to solve a given problem[1,2].

As noted above, intuitively it is easier to add two *n* digit numbers than to multiply them. The statement is based on particular algorithms for computing addition and multiplication. It is clear that the addition of two numbers cannot be performed in a number of steps smaller than *n*. Therefore, we may conclude that the complexity of addition is $O(n)$. On the other hand, in 1971 Schönhage and Strassen discovered an algorithm that requires $O(n \log n \log \log n)$ steps to carry out the multiplication of two *n*-digit numbers on a Turning machine[1]. In fact, the best algorithm's time complexity is used to evaluate the problem's complexity here. However, the existence of better algorithms for computing multiplication cannot be excluded. So it's not reliable to judge which problem is easier based only on the algorithm already been discovered.

Theorist can now classify computational problems into broad categories. **P** problems are those, broadly speaking, that can be solved quickly, such as alphabetizing a list of names. **NP** problems are much tougher to solve but relatively easy to check once you've reached an answer. An example is the traveling salesman problem, finding the shortest possible route through a series of locations. The **P** versus **NP** problem is the single most important question in all of theoretical computer science and one of the most important in all of mathematics[3]. The answer that can **P** from **NP** be separated is still open but most computer scientists quickly came to believe **P != NP** because of the existence of a special **NP** problem: **NP**-complete (**NPC**)problems. The theory of NP-completeness tells us that an efficient solution to any **NPC** problem would imply **P = NP** and an efficient solution to every **NPC** problem. Although a number of ways have been tried but they all failed to prove **P!= NP**. There is little reason to believe that a proof separating **P** from **NP** will be found in the near future [3].

In thousands of different **NPC** problems, the k-satisfiability problem (k-SAT, K>=3) is no doubt the most remarkable one. The k-SAT problem asks whether one can satisfy simultaneously a set of *M* constraints between *N* Boolean variables, where each constraint is a clause built as the logical OR involving *K* variables (or their negations). The interesting threshold phenomena have been observed in k-SAT problems with random generated Boolean formula[4,5]. When the number of variable *N* and of clauses *M* both increase at a fixed value of $\alpha = M / N$, random k-SAT problems become generally SAT at small $\alpha$ and generally UNSAT at large $\alpha$. The existence of a SAT-UNSAT phase transition in the infinite *N* limit has been established rigorously for any *K*, but the critical value $\alpha_c$ (that separate the two phases) has been found only in the (polynomial) K=2 problem where $\alpha_c = 1$. For the NP-complete case $K \geq 3$, as far as we know, the best numerical estimate for $\alpha_c$ at K=3 is 4.26675[6]. The SAT-UNSAT decision problem is also known experimentally to be harder to solve in the neighborhood of $\alpha_c$. It's known that 2-SAT has a continuous phase transition at $\alpha_c$, and that 3-SAT has an abrupt phase transition at $\alpha_c$ [7]. According to phase transitions theory, there must be sudden discontinuities in certain physical properties at the transition. But we still didn't know what physical properties are in the case of 2-SAT and 3-SAT.

Computers are physical systems: the laws of physics dictate what they can and cannot do[8]. Or in other words, information is physical[9]. Since the fundamental law of physics is reversible, many people think it's possible to build a reversible computation without energy consumption[10-12]. The classical computer, which is based on irreversible gates (and gate, or gate), is intrinsically dissipative. In contrast, reversible computer is based on reversible gates (Toffoli gate, Fredkin gate[13]). In principle, there is no energy dissipation in reversible computer. Quantum computer is believed to be a special case of reversible computation. The power of quantum computers is due to the typical quantum phenomena. Several quantum algorithms have been found. In which Grover[14] has shown that quantum mechanics can be useful for solving the problem of searching for a marked item in an unstructured database. In this case, the gain with respect to classical computation is quadratic. But scientists are still trying to figure out what quantum-mechanical properties make quantum computers so powerful and to engineer quantum computers big enough to do something useful. In the same time, even quantum computer is believed not likely to solve NPC problems[3].

As we can see, the time and space complexity analysis now is only about a given algorithm. But many (or even infinite) algorithms may exist for a given problem, how could we know which one is the best or where their time complexity lower bound is. Despite hard work by some of the field's most gifted researchers for over fifty years.

With very few and limited exceptions, lower bounds are still largely in the realm of conjecture[15]. **P** versus **NP** is such kind of problem. It's generally believed that there does not exist algorithm with polynomial time complexity for **NPC** problems. In other words, the time complexity lower bound is supposed to be exponentially for them. It's obvious we could not follow the traditional ideas in order to solve the above problems. We should not analyze only the complexity of a specific algorithm, but instead analyze the problems' complexity. First we should be able to define a problems' complexity and second be able to deduce the lower bound of this complexity. Further on, in theoretical we need to analyze all kind of computer (pervasive computer, not limited to classical electronic computer, but quantum computation, DNA computation etc.). This article will do some research in this field from the energy perspective, using thermodynamics and statistical mechanics as tools, propose a strict definition of problem complexity and investigate mainly several classical problems' complexity.

## Ⅱ.ENERGY AND INFORMATION

Discussion on the relationship between energy and information can be traced back to the Maxwell demon which is proposed by Maxwell in 1867. Maxwell conceived a thought experiment as a way of furthering the understanding of the second law[16]. A Maxwell demon is a construct that can distinguish the velocities of individual gas molecules and then separate hot and cold molecules into two domains of a container, after that the two domains will have different temperature. The result seems to contradict the second law of thermodynamics. Maxwell demon can be expressed by entropy. Entropy is the most influential concept to arise from statistical mechanics. It has three related interpretations[7]. Entropy measures the disorder in a system; Entropy measures our ignorance about a system; Entropy measures the irreversible changes in a system. The definition of entropy is: $S = k \cdot ln W$. In which $k$ is Boltzmann constant, $W$ is the number of all possible microstate which give the same macrostate. Indeed, one can view this thought experiment as giving a fundamental limit on demon efficiency, putting a lower bound on how much entropy an intelligent being must create in order to engage in this kind of sorting process[1,16].

Landauer's principle, first proposed in 1961 by Landauer[17], is a physical principle pertaining to the lower theoretical limit of energy consumption of a computation. It holds that "any logically irreversible manipulation of information, such as the erasure of a bit or the merging of two computation paths, must be accompanied by a corresponding entropy increase in non-information bearing degrees of freedom of the information processing apparatus or its environment".

Landauer's principle asserts that there is a minimum possible amount of energy required to change one bit of information, known as the Landauer limit: $kT ln 2$. $T$ is the temperature of the circuit in kelvins, and ln2 is the natural logarithm of 2. The Landauer's principle will be explained using Maxwell demon. But Bennett[12] pointed out that this former limit was wrong because it is not necessary to use irreversible primitives. Calculation can be done with reversible machines containing only reversible primitives. If this is done, the minimum free energy required is independent of the complexity or number of logical steps in the calculation.

## Ⅲ.THREE CLASSICAL PROBLEMS

From the view of Landauer principle, owing to the fact that pervasive computing process can be described as the thermodynamic cycle process. The idea of Landauer principle was extended through mapping three classic problems (sorting, ordered searching and max of *N* numbers) into Maxwell demon thought experiment. Then the energy consumption lower bounds of all possible algorithms were rigorous deduced.

**A. Search for the max value in unordered *N* numbers**

Suppose we have *N* nearly independent particles (i.e. ideal gas, there were no interactions between particles) in a container with equally left and right domains. In the beginning these *N* particles are randomly distributed in two domains. Then the demon began to separate the particle with fastest velocity into the left domain and the others into the right domain. If these *N* particles were treated as *N* variables, and their velocities as the corresponding variable's values, through this mapping the demon can be treated as a special purpose computer that deals with the problem of searching for the max value in unordered *N* numbers.

According to the second law of thermodynamics, the demon must consume energy in order to accomplish this task. The minimum energy required is $T(S_1 - S_2)$, in which $S_2$ is the entropy after separation while $S_1$ is the entropy before.

If there is only one particle, the probabilities that particle belongs to the two domains are both 1/2, the corresponding entropy is:

$$S_1 = k\frac{1}{2}\ln 2 + k\frac{1}{2}\ln 2 = k\ln 2$$

After separation, suppose the particle was limited to the left domain. The probability that particle belongs to the left domain is 1 and the probability belongs to the right domain is 0. The corresponding entropy is

$$S_2 = k1\ln 1 = 0$$

According to the second law of thermodynamics, the minimum energy required by the demon is $T(S_1 - S_2) = kT\ln 2$. This is exactly the Landauer principle. Because these *N* particles are nearly independent and entropy as extensive parameters, the minimum energy required in order to separate *N* particles will be $NkT\ln 2$. In another way, the number of complexion that the *N* particles can distribute is $W_1 = 2^N$, the corresponding entropy is $S_1 = k\ln W_1 = Nk\ln 2$. After separation, the number of complexion is 1 and the corresponding entropy is $S_2 = k\ln W_2 = k\ln 1 = 0$.

There is still one big problem unsettled in the above problem mapping. In the above problem, the values of these *N* variables are uniform distributed. If we map these values direct into the particle's velocity, these *N* particles will be in nonequilibrium state with no corresponding temperature. In equilibrium state the velocity of particles should be maxwellian distribution. In order to avoid the puzzle of temperature, we should make a mapping between the uniform distribution and maxwellian distribution.

**B. Sorting of *N* numbers**

The container in Maxwell demon was divided into *N* equal parts. The demon then separates these *N* particles into the *N* left-to-right positions according to their velocities. The entropy before separation is:

$$S_1 = k\ln W_1 = k\ln N^N = kN\ln N$$

The entropy after separation is 0 and the minimum energy required to solve this problem was $kTN\ln N$.

**C. Searching in *N* sorted numbers**

The container in Maxwell demon was divided into *N* equal parts. The demon places the specific particle in the right place according to its velocity. The entropy before separation is:

$$S_1 = k\ln W_1 = k\ln N$$

The entropy after separation is 0 and the minimum energy required to solve this problem was $kT \ln N$.

**D. Definition of problem complexity**

Based on the preceding analysis of three classical problem, as we can see, the essential part of solving problem by any kind of computers(plus algorithm) is an entropy reduction (from the disorder input to the orderly output) process. This entropy reduction could be defined as the problem's complexity. The energy consumed by any kind of computer which needs to solve the problem can not less than this lower bound.

## IV. REAL ALGORITHM IN CLASSICAL COMPUTER

With regard to any specific algorithms in classical computer, it's always composed by three basic structures: sequence, selection and circulation. The time complexity of an algorithm is generally connected with cycle index. The basic operations a computer can perform in one cycle are nothing but add, multiplication, assignment and logical comparison, etc. We'll first analyze the energy consumption of these basic operators, after that the specific algorithm will be analyzed.

**A. Basic operators**

As for the entropy reduction of one bit ">" operator, the problem space is 3 bit. There are eight possible initial states, and in thermal equilibrium they will occur with equal probability. How much entropy reduction will occur in a machine cycle? The initial and final machine states are shown in Table 1. State A, B, C and D occur with a probability of 1/4 each.

Table 1. One bit ">" operator which maps eight possible states onto only four different states

| Before cycle | | | After cycle | | | Final |
|---|---|---|---|---|---|---|
| bit | bit | result | bit | bit | result | state |
| 0 | 0 | 0 | 0 | 0 | 0 | A |
| 0 | 0 | 1 | 0 | 0 | 0 | A |
| 0 | 1 | 0 | 0 | 1 | 0 | B |
| 0 | 1 | 1 | 0 | 1 | 0 | B |
| 1 | 0 | 0 | 1 | 0 | 1 | C |
| 1 | 0 | 1 | 1 | 0 | 1 | C |
| 1 | 1 | 0 | 1 | 1 | 0 | D |
| 1 | 1 | 1 | 1 | 1 | 0 | D |

The initial entropy was

$$S_i = k \ln W = -k \sum \rho \ln \rho = -k \sum \frac{1}{8} \ln \frac{1}{8} = 3k \ln 2$$

The final entropy was

$$S_f = -k \sum \rho \ln \rho = -k \cdot 4 \cdot (\frac{1}{4} \ln \frac{1}{4}) = 2k \ln 2$$

The difference $S_1 - S_2 = k \ln 2$

If the ">" operator was tested against two *M* bit string, the entropy reduction will still be $k \ln 2$. To our surprise, the complexity of ">" operator has nothing to do with the length of string been tested. In the same time, just like Landauer principle, our method of reasoning gives no guarantee that this minimum is in fact achievable. It should be mentioned that the algorithm with the above minimum energy consumption is still not available now. For classical computer we may never find such algorithm(because it has to do the operation one bit by one bit) although we believe in the possibility in other kind of computers. For example, suppose we have two parallel lights with different frequencies, their frequencies as the numbers to be compared, the ">" operator can be easily done with a glass prism. This result can be easily expanded to other basic operators: "<","<=",">=","=","!=". Their complexities are all $k\,ln\,2$ that have nothing to do with the length of string been compared.

In fact, we might found a flaw in the theory of reversible computation. Any irreversible function can be embedded into a reversible function. For example, in order to construct the NAND gate (irreversible) with the Toffoli gate(reversible, Table 2)(*1*), we could set c=1, so that c'=0 if and only if a=1 and b=1, that is $c' = 1 \oplus ab = a\text{NAND}b$. But how can we set c=1? Does it dissipate at least $kT\,ln\,2$ heat according to Landauer principle? That is to say, in order to use this reversible NAND gate, we need first dissipate heat which is irreversible. Then how can we promise the whole process is reversible?

Table 2. The truth table for the Toffoli gate

| a | b | c | a' | b' | c' |
|---|---|---|----|----|----|
| 0 | 0 | 0 | 0  | 0  | 0  |
| 0 | 0 | 1 | 0  | 0  | 1  |
| 0 | 1 | 0 | 0  | 1  | 0  |
| 0 | 1 | 1 | 0  | 1  | 1  |
| 1 | 0 | 0 | 1  | 0  | 0  |
| 1 | 0 | 1 | 1  | 0  | 1  |
| 1 | 1 | 0 | 1  | 1  | 1  |
| 1 | 1 | 1 | 1  | 1  | 0  |

As for the basic arithmetic operators, the complexity of add and multiplication can be easily defined. Take the add operation of two *n* bit number into account.
The problem space is $n + n + (n+1) = 3n+1$ bit. The initial entropy was

$$S_1 = k\,ln\,W = k\,ln\,2^{3n+1} = (3n+1)k\,ln\,2$$

The final entropy was

$$S_2 = k\,ln\,W = k\,ln\,2^{2n} = 2nk\,ln\,2$$

The difference $S_1 - S_2 = (n+1)k\,ln\,2$

Take the multiplication operation of two n bit number into account.
The problem space is $n + n + 2n = 4n$ bit. The initial entropy was

$$S_1 = k\,ln\,W = k\,ln\,2^{4n} = 4nk\,ln\,2$$

The final entropy was

$$S_2 = k\,ln\,W = k\,ln\,2^{2n} = 2nk\,ln\,2$$

The difference is $S_1 - S_2 = 2nk\,ln\,2$

It's clear that the complexity of add and multiplication are in the same level (both $O(n)$). So the algorithm discovered by Schönhage and Strassen is certainly not the best in theory. Again, our method of reasoning gives no guarantee that this minimum is in fact achievable.

**B. Specific algorithm of three problems**

In the specific algorithm that is used to solve the above three problems, the main structure is usually circulation. We'll use the sort problem as example in the following discussion. Before sort the N numbers(m bit) has to be initialized(from uncertainty state to certainty state) that need $kT\,ln\,2 \cdot m$ energy according to Landauer principle, but it does not belong to energy consumed by sort process. The following statements (in C language) or like are expected to execute in one cycle of comparison sort algorithm.

…

```
if (x[i]>x[j])                  1
    {
        temp=x[i];              2
        x[i]=x[j];              3
        x[j]=temp;              4
    }
```
…

According to Landauer principle, the three assignment operation (determinacy operation, from one certainty state to another certainty state, sentences 2,3,4) need not consume energy in theory. The one need energy to operate is the branch statement (sentence 1 which is a kind of uncertainty operation). The operation it performs is ">" with the lower bound of energy consumption as $kT\ln 2$. According to the previous derivations, the three classical problems' complexities are $Nk\ln 2$, $kN\ln N$, $k\ln N$ respectively. Then the time complexity lower bounds of all algorithms are respectively

$$\frac{Nk\ln 2}{k\ln 2} = N, \quad \frac{kN\ln N}{k\ln 2} = N\log_2 N, \quad \frac{k\ln N}{k\ln 2} = \log_2 N.$$

The time complexity of the best algorithms(For sort, it refers to comparison sort) discovered to solve the three problems are $O(N)$, $O(N\log_2 N)$ and $O(\log_2 N)$ respectively. Their corresponding energy consumptions are exactly the same with the lower bounds deduced. It seems the best algorithms (with regard to classical computer) for them have all been discovered.

As for the radix sort which is non-comparative integer sorting algorithm, the time complexity is $O(N)$, but it needs extra data structure called bucket(at least N buckets) compared with comparison sort, each bucket needs at least $\log_2 N$ bit. In each of $N$ cycle, the radix sort need to write one bucket, this will cost $kT\ln 2 \cdot \log_2 N = kT\ln N$ energy according to Landauer principle. Even radix sort cannot exceed the above lower bound.

**C. Rethinking of three classical problems**

With entropy reduce as the definition of problem's complexity, we can rethink three classical problem from another point of view.

In order to search the maximum value of $N$ numbers, we need first to define the problem space as the follows.
10101010   10000100   00001111   1 0 0
The foregoing part consists of $N$ numbers, each number has $M$ bit. The posterior part consists of $N$ bits. These $N$ bits have and only have one bit set to 1, which represents the corresponding number is the max one.
The initial entropy was
$S_1 = k\ln W = k\ln 2^{NM+N} = (NM+N)k\ln 2$
The final entropy was
$S_2 = k\ln W = k\ln 2^{NM} = NMk\ln 2$
The difference   $S_1 - S_2 = Nk\ln 2$

As for the sorting problem of $N$ numbers, the problem space can be defined as the follows.
   10101010   10000100   00001111   00101011   11  10  00  01
The foregoing part consist of $N$ numbers, each number has $M$ bit. The posterior part consists of $N\,log_2\,N$ bits. The first $log_2\,N$ represent the first number's position after sorting.
The initial entropy was
$$S_1 = k\,lnW = k\,ln\,2^{NM+N\,log_2\,N} = (NM + N\,log_2\,N)kln2$$
The final entropy was
$$S_2 = k\,lnW = k\,ln\,2^{NM} = NMkln2$$
The difference $\quad S_1 - S_2 = N\,log_2\,Nk\,ln\,2 = kN\,ln\,N$

As for the problem of searching in $N$ ordered numbers, the problem space can be defined as the follows.
   10101010   10000100   00101011   00001111   00101011   10

The foregoing part consists of $N$ ordered numbers plus the number to be searched, each number has $M$ bit. The posterior part consists of $log_2\,N$ bits that represent the position after searching.
The initial entropy was
$$S_1 = k\,lnW = k\,ln\,2^{NM+M+log_2\,N} = (NM + M + log_2\,N)kln2$$
The final entropy was
$$S_2 = k\,lnW = k\,ln\,2^{NM+M} = (NM + M)kln2$$
The difference $\quad S_1 - S_2 = log_2\,Nk\,ln\,2 = k\,ln\,N$
These results are all in coincidence with the previous analysis.

**D. Two $n \times n$ matrix multiplication problem**

The famous two $n \times n$ matrix multiplication problem. For a long time it was assumed that one need $n^3$ operations to multiply two $n \times n$ matrices. But in 1969, Volker Strassen showed that the problem can be solved by an ingenious recursive algorithm in $O(n^{2.81})$ operations. Over the past forty years, this exponent has undergone a sequence of improvements, and now stands below 2.4[15]. In this problem, suppose one element in the two matrix is $m$ bit, after multiplication, an element in the result matrix will be $2m+log_2\,n$ ($\sum_n m\,bit \times m\,bit$) bit in order to store the full information. The whole result matrix will consist of $(2m+log_2\,n)n^2$ bit. The complexity lower bound of this problem is $O(n^2(log_2\,n+2m))$.

## V. CONCLUSIONS AND DISCUSSIONS

In this paper entropy reduction was proposed as the strict definition of problem complexity. It's based on energy and can be applied to any kind of computation, behind which is the second law of thermodynamics. The preliminary results brought us fresh and profound understanding of computation complexity. It's quite natural to provoke further thoughts in three fields.

Is reversible computation really possible? We really don't think reversible computation is possible. From the preceding analysis, it's clear that the demon must dissipate certain heat into surroundings in order to solve these problems. The whole process is certainly irreversible. In fact, we think the situation here is somewhat like the debate on microscopic reversibility and macroscopic irreversibility in physics. Although the quantum computer is considered to be composed of reversible gates (microscopic reversibility), the whole process of solving problem will still be irreversible (macroscopic irreversibility).

Could computer utilize energy with super-linear capacity? With the previous study we can get an important inspiration: Because the energy consumed by classical computer is linear with time, has nothing to do with the scale of the problem to be solved. Those computers that can utilize energy with super-linear capacity are what we should seek. We believe in this possibility although we did not see any research in this field. In fact, quantum computation may be such kind of computer. Grover has shown that quantum mechanics can be useful for solving the problem of searching for a marked item in an unstructured database (the problem's complexity is $O(N)$ level).

The number of required iterations in Grover's algorithm is $O(\sqrt{N})$. In this case, the gain with respect to classical computation is quadratic. If the energy cost is still suitable in this situation, we have reason to believe that the quantum computation has the ability to pay $O(N)/O(\sqrt{N})=O(\sqrt{N})$ level energy(super-linear)in one iteration. Could this capacity be the real power behind quantum computation?

A conjecture about NPC problem. Is there any possibility that certain problem exists which need exponential level energy to solve? We propose the following conjecture.

The energy cost needed by **NPC** problems is exponential with the scale of the problem.

If this conjecture is proved, with regard to the classical computation we can get the inference that **NP!=P**. The reason is quite simple because the classical computer can only consume energy linearly (the power of CPU is constant, and the basic operations it can deal are all linearly). It's impossible to find any algorithms that can solve this kind of problem with **P** level time complexity. The result is the same for quantum computation unless it can utilize energy with exponential capacity.